\documentstyle[12pt,epsf]{article}
\renewcommand{\theequation}{\thesection.\arabic{equation}}

\newcommand{\tu}[1]{T^u_{#1}}
\newcommand{\td}[1]{T^d_{#1}}
\newcommand{\te}[1]{T^e_{#1}}
\newcommand{\st}[1]{\tilde{s}_{#1}}
\makeatletter
\def\lsim{\mathrel{\mathpalette\@versim<}}
\def\@versim#1#2{\vcenter{\offinterlineskip
        \ialign{$\m@th#1\hfil##\hfil$\crcr#2\crcr\sim\crcr }}}
\makeatother

\begin{document}
\begin{titlepage}
  \begin{flushright}
    KUNS-1619\\[-1mm]
    hep-ph/0003220
  \end{flushright}
  \begin{center}
    \vspace*{1.4cm}
    
  {\Large\bf Mass Matrices in $E_6$ Unification} 
  \vspace{1cm}
  
  M.~Bando\footnote{E-mail address: bando@aichi-u.ac.jp},
  T.~Kugo\footnote{E-mail address: kugo@gauge.scphys.kyoto-u.ac.jp} 
  and 
  K.~Yoshioka\footnote{E-mail address:
    yoshioka@gauge.scphys.kyoto-u.ac.jp}
  \vspace{5mm}
  
  $^*$ {\it Aichi University, Aichi 470-0296, Japan}\\
  $^{\dagger,\ddagger}$ {\it Department of Physics, Kyoto University
    Kyoto 606-8502, Japan}
  \vspace{1.5cm}
  
  \begin{abstract}
    We study a supersymmetric $E_6$ grand unified model in which the
    $SU(5)$ ${\bf 5}^*$ components are twisted in the third generation
    {\bf 27}. Supplementing the adjoint Higgs field to a model
    analyzed previously, we calculate the mass matrices for the up and
    down quarks and charged leptons. Although the number of free
    parameters is less than that of observables, an overall fitting
    to the observed masses and mixing angles is shown to be
    possible. Most notably, we find two novel, parameter-independent
    relations between the lepton 2-3 mixing angle $\theta_{\mu\tau}$
    and the quark masses and CKM mixing angles that are in good
    agreement with the large lepton mixing recently observed.
  \end{abstract} 
\end{center}
\end{titlepage}
\setcounter{footnote}{0}

\section{Introduction}

The discovery of neutrino oscillation events in Superkamiokande
experiments~\cite{SK} has stimulated us to reconsider the fermion mass
hierarchy problem in unified theories. One of the most mysterious
problems is the remarkable contrast of the mixing structure of leptons 
to that of quarks. Most of us believe in the possible unification of
the existing matter content observed in the low energy region and that 
the realization of a grand unified theory (GUT) is one of the most
challenging problems in particle physics. In this sense, the
experimental data~\cite{SK}-\cite{macro} indicating a large mixing of
the muon neutrino provides us with an important clue in pursuing
unification and many authors have investigated neutrino physics along
this line~\cite{neutrino}.

The existence of right-handed neutrinos suggests a left-right
symmetric gauge group which is beyond the $SU(5)$ group~\cite{su5}, in
which all fermions of one generation are combined into a single
representation {\bf 16}. In this sense, $SO(10)$~\cite{so10} would be
an attractive candidate for the unified gauge group. However, the
characteristic feature of the neutrino mixing structure seems to
suggest that such $SO(10)$ unifications are unfeasible they would give
rise to a complete parallelism between quarks and leptons.

Among the possible simple gauge groups, $E_6$~\cite{e6} is essentially
the unique candidate for the unified gauge group. Indeed, if one
requires the following three conditions for the unified group, 
only $SO(10)$ and $E_6$ remain: (i) all the fermions, including the
right-handed neutrinos of one family, belong to a single irreducible
representation; (ii) the group is automatically anomaly free; and
(iii) the group allows complex representations that contain our
low-energy chiral fermions but not their mirror fermions. If we
further add the condition that (iv) there exists freedom in the
fermion content to avoid the parallelism between the quark and lepton
mass structures, then we are left with only the $E_6$ gauge
group. Most remarkably, one of characteristic features of the $E_6$
group is that we have a freedom in choosing the down quark and charged
lepton components in the fundamental representation {\bf 27}.

Our final goal is to unify all the fermions of three generations, but
before realizing that, we should understand the difference between the
mixing structures of the quark and lepton sectors. This is closely
connected with the origin of the hierarchical structure of fermion
masses. There have been various approaches to these problems, and to
this time, most of them either take {\it ad hoc} employ for
hierarchical Yukawa couplings or assume certain family
symmetries. Among the latter approaches, which we expect to open a
gate for understanding the origin of generation, the simplest one may
be to use a flavor $U(1)$ symmetry~\cite{u1}. In that case, the
smallness of the Yukawa couplings are attributed to the
higher-dimensional interaction terms suppressed by some fundamental
scale $M_P$.

In a previous paper~\cite{BK}, the authors constructed a
supersymmetric $E_6$ unified model with an extra $U(1)$
symmetry. There we showed that E-twisting family structure can
reproduce all the characteristic features of the fermion mass
matrices, not only the quark/lepton Dirac masses but also the neutrino
Majorana masses. Despite the fact that a common $U(1)$ charge is
assigned to all members in a {\bf 27} of each generation, the model
explained the qualitative features of the different mass hierarchies
among generations for up and down quark sectors, as well as the mixing
angles for quarks and for leptons.

In this paper, we study this model more quantitatively and aim to
check whether the resultant predictions are consistent with the
present experimental results. To complete our scenario, we
have to introduce a new Higgs field in order to allow for a difference
between the quark and lepton masses and mixings. A minimal choice is
to introduce a Higgs field of the adjoint 
representation {\bf 78}. This Higgs field is actually necessary also
in order to break the $E_6$ gauge symmetry down to the standard gauge
group $SU(3)\times SU(2)\times U(1)$. This is because there is no
component in the fundamental representation ({\bf 27}) that is
non-singlet under the $SU(5)$ symmetry but singlet under the standard
gauge group. In fact, with a particular assumption for the Higgs
potential, the adjoint representation Higgs can cause the desired
symmetry breaking even with the doublet-triplet
splitting~\cite{break}. The newly introduced Higgs field also induces
effective Yukawa couplings which come from the higher-dimensional
interaction terms via the Froggatt-Nielsen mechanism~\cite{fn}. It is
found that such (non-leading) contributions can actually lead to
differences between the quarks and leptons as well as between the up
and down quarks. We analyze the masses and mixing angles by taking
account of this additional Higgs field and find several interesting
parameter-independent relations among the experimental observables,
including a large lepton mixing angle.

This paper is organized as follows. In section 2, we first present the 
field content and the charge assignments in our model with E-twisting
family structure by adding a Higgs field {\bf 78}. We also explicitly
list the induced higher-dimensional terms involving the {\bf 78}
field. The structure of the additional couplings is analyzed in
section 3. We show how this contribution affects the quark and lepton
Yukawa couplings in the following two sections. In section 6 we carry
out both the analytic and numerical analyses to compare the results
with the present experimental data. In spite of the complicated forms
of the Yukawa matrices, we can find interesting relations between the
quark and lepton mixing angles. In section 7 we give summarizing
discussion and further comments. The Appendix is devoted to describing
the diagonalization of the $3\times3$ mass matrices of the type
appearing in the text.

\section{Model}
\setcounter{equation}{0}

The supersymmetric $E_6$ unification model that was considered in
Ref.~\cite{BK} contains, in addition to an $E_6$ gauge vector
multiplet, chiral matter multiplets corresponding to the three
generation fermions ($\Psi_i$ $(i=1,2,3)$) and two pairs of Higgs
fields ($H$, $\bar H$) and ($\Phi$, $\bar\Phi$). The former Higgs,
$H$, is for the electroweak symmetry breaking (also for the fermion
masses) and the other, $\Phi$, is responsible for realizing the
E-twisting (generation flipping) structure. In this paper, we
additionally introduce a chiral Higgs multiplet $\phi({\bf 78})$,
which is necessary to break the GUT to the standard gauge group. In
Table 1, we summarize all the fields we need in this paper.
\begin{table}[htbp]
\begin{center}
\begin{tabular}{c|c|c|c|c|c|c|c|c|c}\hline
  & $\Psi_1$ & $\Psi_2$ & $\Psi_3$ & $H$ & $\bar H$ & $\Phi$ &
  $\bar\Phi$ & $\phi$ & $\Theta$ \\ \hline 
  $E_6$ & {\bf 27} & {\bf 27} & {\bf 27} & {\bf 27} 
  & ${\bf 27^*}$ & {\bf 27} & ${\bf 27^*}$ & {\bf 78} & {\bf 1} \\
  \hline
  $U(1)$ charge & 3 & 2 & 0 & 0 & 0 & $-4$ & 4 & $-2$ & $-1$ \\
  \hline
  $R$ parity & $-$ & $-$ & $-$ & + & + & + & + & + & + \\ \hline
\end{tabular}
\caption{$E_6$ representations and $U(1)$ charge assignment.}
\end{center}
\end{table}
The $E_6$ singlet field $\Theta$ with $U(1)$ charge $-1$ plays the
important role that its suitable powers compensate for the mismatch of
the $U(1)$ charge in the superpotential interaction terms. The $U(1)$
flavor symmetry discriminates generations and induces a hierarchy
among them. It should be noted that all the quarks and leptons in one
generation have a common $U(1)$ quantum numbers.

Since we are interested in the mass terms of the ordinary fermions as
well as superheavy fermions of the GUT scale, we list all the
superpotentials which are invariant under $R$ parity, $U(1)$ and $E_6$
and give masses of matter superfields $\Psi_i({\bf 27})$. The Yukawa
interactions are given by
\begin{eqnarray}
  W_Y(H) &=& y_{ij}\,\Psi_i({\bf 27})\Psi_j({\bf 27})H({\bf 27})
  \left({\Theta\over M_P}\right)^{f_i+f_j}, \nonumber \\
  W_Y(\Phi) &=& y_{ij}'\,\Psi_i({\bf 27})\Psi_j({\bf 27}) 
  \Phi({\bf 27}) 
  \left({\Theta\over M_P}\right)^{f_i+f_j-4},
  \label{yukawa}
\end{eqnarray}
where $f_i$ denotes the $U(1)$ charge of the $i$-th generation. The
coupling constants $y$ and $y'$ are assumed to be order 1. However,
the effective Yukawa coupling constants become multiplied by
additional powers of $\lambda=\langle\Theta\rangle/M_P$ coming, from
the powers of $\Theta$ required by $U(1)$ quantum number matching (the
Froggatt-Nielsen mechanism). We assume throughout this work that
$\lambda$ is of the order of the Cabibbo angle
$\sim0.22$~\cite{u1,anomU1}. We also suppose that only the $SU(2)$
doublet components of $H$ can have the electroweak scale vacuum
expectation value (VEV).

With the adjoint Higgs field $\phi$ having $U(1)$ charge $-2$, there
is the following higher-dimensional operator which eventually give
small mass terms:
\begin{eqnarray}
  W_{\phi} &=& z_{ij} M_P^{-1}\phi({\bf 78})\Psi_i({\bf 27})
  \Psi_j({\bf 27})H({\bf 27}) 
  \left({\Theta\over M_P}\right)^{f_i+f_j-2}.
  \label{Zyukawa}
\end{eqnarray}
Here, $z_{ij}$ are order 1 couplings. Precisely speaking, there are
two ways for contracting the $E_6$ group indices in this
superpotential, which we will discuss in detail in the next
section. There also exist the following higher-dimensional operators,
which give rise to the right-handed neutrino Majorana masses:
\begin{eqnarray}
  W_R &=& x_{ij}M_P^{-1}\Psi_i({\bf 27})\Psi_j({\bf 27})
  X_k(\overline{\bf 27})X_l(\overline{\bf 27}) \left({\Theta\over M_P} 
  \right)^{f_i+f_j+{\hspace{1pt}x}_k+{\hspace{1pt}x}_l}.
\end{eqnarray}
Here $X_i$ represents $\Phi(\overline{\bf 27})$ 
or $H(\overline{\bf 27})$ with $U(1)$ charge $x_i$, and the couplings
$x_{ij}$ are of order 1. For the resulting mass texture of the
right-handed neutrinos, see Ref.~\cite{BK}.

For later convenience, we name the component fields 
of $\Psi({\bf 27})$ as follows. The representation {\bf 27} is
decomposed under $SO(10)\subset E_6$ as
\begin{eqnarray}
  {\bf 27} &=& {\bf 16} + {\bf 10} + {\bf 1}, 
\end{eqnarray}
which are further decomposed under $SU(5)\subset SO(10)$ as
\begin{eqnarray}
  \begin{array}{ccccccc}
    {\bf 16} &=& {\bf 10} &+& {\bf 5}^* &+& {\bf 1},\\[2mm]
    && \left[u^{c\,i},\pmatrix{u_i\cr d_i\cr}, e^c\right] 
    && (d^{c\,i},e,-\nu) && \nu^c \\
  \end{array} \nonumber
\end{eqnarray}
\begin{eqnarray}
  \begin{array}{ccccccccc}
    {\bf 10} &=& {\bf 5} &+& {\bf 5}^*,
    &\qquad \quad & {\bf 1} &=& {\bf 1}. \\[2mm]
    && (D_i, E^c, -N^c) && (D^{ci},E,-N) 
    &\qquad \quad & && S  \\
  \end{array}
  \label{27}
\end{eqnarray}
An interesting fact is that ${\bf 5}^*$ appears twice in 
each ${\bf 27}$, i.e., ${\bf 5}^*$ of ${\bf 16}$ ((${\bf 16,5^*}$))
and ${\bf 5}^*$ of ${\bf 10}$ ((${\bf 10,5^*}$)) which we refer to as
the `E-parity' doublet. It is due to this doubling that we have the
freedom to choose the low-energy ${\bf 5}^*$ candidates. This actually
implies that the embedding of $SO(10)$ into $E_6$, such that
$SU(5)_{\rm GG}\subset SO(10)\subset E_6$ with Georgi-Glashow
$SU(5)_{\rm GG}$, possesses a freedom of rotation of $SU(2)_R$. The
doubling of ${\bf 5}^*$ in each ${\bf 27}$ also provides the
low-energy surviving down-type Higgs field with the freedom of a
mixing parameter between the two ${\bf 5}^*$ representations in
$H({\bf 27})$:
\begin{eqnarray}
  H({\bf 5}^*) &=& H({\bf 10,5^*})\cos\theta+H({\bf 16,5^*})
  \sin\theta.
\end{eqnarray}

Let us pick up the low-energy matter fields among the 
three $\Psi_i({\bf 27})$ of the above. The up-quark sector is unique,
since ${\bf 10}$ and ${\bf 5}$ of $SU(5)$ appear only once in each
${\bf 27}$. As for the three families of (right-handed) down quarks,
there is a freedom in choosing three from the six ${\bf 5}^*$
representations in the three $\Psi_i({\bf 27})$. We have classified
possible typical scenarios in Ref.~\cite{BK}: (i) parallel family
structure; (ii) non-parallel family structure; and (iii) E-twisted
structure. Among these three possibilities, we here investigate the
simplest and probably most attractive option, namely the E-twisted
structure:
\begin{eqnarray}
  ({\bf 5}^*_1,\ {\bf 5}^*_2,\ {\bf 5}^*_3) &=& 
  \bigl(\Psi_1({\bf 16,5^*}),\ \Psi_2({\bf 16,5^*}),\ 
  \Psi_3({\bf 10,5^*})\bigr).
  \label{twist}
\end{eqnarray}
This structure implies that the third family ${\bf 5}^*$ falls 
into ${\bf 10}$ of $SO(10)$, which is E-twisted from that of the other
two families. This twisting is realized by the suitable VEVs of the
Higgs field $\Phi$ and the usual Higgs $H$, as shown in Ref.~\cite{BK}
and explained briefly below.

In order to get the actual forms of mass matrices, we should know the
couplings containing the neutral components of Higgs fields, 
$H({\bf 1})$, $H({\bf 16,1})$, $H({\bf 16,5}^*)$, $H({\bf 10,5^*})$
and $H({\bf 10,5})$, which correspond to the components $S$, $\nu^c$,
$-\nu$, $-N$ and $-N^c$, respectively, of $\Psi({\bf 27})$ in
Eq.~(\ref{27}). In addition to this, we need also write down the form
of the coupling (\ref{Zyukawa}) at the component level. The full
effective mass matrices for the quarks and leptons are the sums of the
contributions from direct Yukawa couplings (\ref{yukawa}) and the
induced ones from the higher-dimensional couplings
(\ref{Zyukawa}). Since the explicit forms of the tree-level Yukawa
couplings have been presented in Ref.~\cite{BK}, we here present those
of Eq.~(\ref{Zyukawa}) in the next section.

\section{Coupling of $\phi({\bf 78})$}
\setcounter{equation}{0}

In this section we present the precise form of the higher-dimensional
interaction (\ref{Zyukawa}) containing the adjoint 
Higgs $\phi({\bf 78})$ and give the explicit component expressions for
the effective Yukawa couplings induced by the VEVs 
of $\phi({\bf 78})$. 

The $E_6$ adjoint $\phi({\bf 78})$ is decomposed as follows under the
subgroup $SU(3)_L\times SU(3)_R\times SU(3)_C\subset E_6$: 
\begin{equation}
  {\bf 78} = {\bf 8}_L + {\bf 8}_R + {\bf 8}_C +({\bf 3,3,3}) 
  +({\bf 3^*,3^*,3^*}).
\end{equation}
Then, ${\bf 8}_{R(L)}$ is further decomposed 
under $SU(2)_{R(L)}\subset SU(3)_{R(L)}$ into
\begin{equation}
  {\bf 8}_{R(L)} = {\bf 3}_{R(L)} + {\bf 2}_{R(L)} + {\bf 2}_{R(L)}^*
  + {\bf 1}_{R(L)}, 
\end{equation}
among which the components of ${\bf 8}_R$ and the $SU(2)_L$ singlet
part of ${\bf 8}_L$ are neutral under the standard gauge symmetry. If
$SU(2)_R$ is broken by the VEV of the neutral components 
of ${\bf 3}_R$, ${\bf 2}_R$ and/or ${\bf 2}_R^*$, this can create
differences between the up and down quark sectors. However, it turns
out that ${\bf 2}_R$ and ${\bf 2}_R^*$ do not yield sufficiently large 
differences. Thus we here assume that only the third component
$\phi_{R3}$ of
\begin{equation}
  \phi({\bf3}_R) = \sum_{a=1}^3 \phi_{Ra} {\tau^a\over\sqrt2}
  ={1\over\sqrt2}\pmatrix{\phi_{R3} & \phi_{R1}-i\phi_{R2} \cr 
    \phi_{R1}+i\phi_{R2}& -\phi_{R3} \cr}
\end{equation}
develops a non-vanishing VEV $\omega$, normalized by $\lambda^2M_P$
for later convenience (see (\ref{78VEV})). Since we also wish to have
differences between the down quarks and charged leptons (especially in
the second generation), we assume that the $SU(2)_R$ singlet 
${\bf 1}_R$ in ${\bf 8}_R$ and the $SU(2)_L$ singlet ${\bf 1}_L$ in
${\bf 8}_L$ develop non-vanishing VEVs $\chi_R$ and $\chi_L$
(normalized by $\lambda^2M_P$), respectively. Retaining only these
three VEVs, $\chi_R$, $\chi_L$ and $\omega$, we have
\begin{eqnarray}
  \frac{\langle\phi({\bf 8}_R)\rangle}{M_P} &=& \lambda^2
  \pmatrix{\omega+\chi_R & 0 & 0 \cr 0 & -\omega+\chi_R & 0 \cr 0 & 0
    & -2\chi_R}, \nonumber \\ 
  \frac{\langle\phi({\bf 8}_L)\rangle}{M_P} &=&
  \lambda^2\pmatrix{\chi_L & 0 & 0 \cr 
    0 & \chi_L & 0 \cr 0 & 0 & -2\chi_L}.
  \label{78VEV}
\end{eqnarray}
(For a proper normalization of generators, $\omega$ and $\chi$ should
be replaced by $\omega/\sqrt2$ and $\chi/(2\sqrt3)$.) \ Now we derive
the effective Yukawa couplings resulting from the higher-dimensional
interaction (\ref{Zyukawa}), which actually reads as the following two
independent $E_6$-invariant terms:
\begin{eqnarray}
  W_\phi&=& \sum_{i,j} s_{ij}M_P^{-1} \Psi_i({\bf 27})
  \Psi_j({\bf 27}) \big(\phi({\bf 78})H({\bf 27})\big)_{\bf 27}
  \left({\Theta\over M_P}\right)^{f_i+f_j-2} \nonumber \\
  && +\sum_{i,j} a_{ij}M_P^{-1} \big(\phi({\bf 78}) 
  \Psi_i({\bf 27})\big)_{\bf 27} \Psi_j({\bf 27})H({\bf 27}) 
  \left({\Theta\over M_P}\right)^{f_i+f_j-2}.
  \label{phi}
\end{eqnarray}
The $O(1)$ coupling constant $s_{ij}$ in the first term is clearly
symmetric under the exchange $i\leftrightarrow j$, by definition. The
second term coupling $a_{ij}$, on the other hand, need not have
definite symmetry. However, we can always redefine it to become
anti-symmetric through the following procedure. Since ${\bf 78}$ is
equivalent to the generator of the $E_6$ group, we have the identity
\begin{eqnarray}
  \big(\phi({\bf 78})\Psi_i({\bf 27})\big)_{\bf 27} \Psi_j({\bf 27})
  H({\bf 27}) +\Psi_i({\bf 27})\big(\phi({\bf 78}) 
  \Psi_j({\bf 27})\big)_{\bf 27} H({\bf 27}) && \nonumber \\[1mm]
  +\Psi_i({\bf 27})\Psi_j({\bf 27})\big(\phi({\bf 78}) 
  H({\bf 27})\big)_{\bf 27} &=& 0.
\end{eqnarray}
This shows that the symmetric part of the second term 
coupling $a_{ij}$ in Eq.~(\ref{phi}) is equivalent to the first term
coupling $s_{ij}$. We can therefore assume that the second coupling 
constant $a_{ij}$ is anti-symmetric without loss of generality by a
redefinition of $s_{ij}$.

The above two types of couplings in Eq.~(\ref{phi}) in fact correspond to
the two independent couplings through the representations {\bf 27} and
{\bf 351} contained in the product $\bf 78\times 27=27+351+1728$,
which are contracted with their complex conjugate representations in 
${\bf 27\times 27}=\overline{\bf 27} +\overline{\bf 351}_a
+\overline{\bf 351}'_s$. We can thus see that the second term in
Eq.~(\ref{phi}), with antisymmetric coupling constant $a_{ij}$, is
equivalent to the antisymmetric coupling 
$\big(\Psi_i({\bf 27})\Psi_j({\bf 27})\big)_{\overline{\bf 351}_a}
\big(\phi({\bf 78})H({\bf 27})\big)_{\bf 351}$. 
However, writing this coupling as in Eq.~(\ref{phi}) is more
convenient. This is because the action of 
$\phi({\bf 78})$ on $F({\bf 27})$ ($F=\Psi,H$)
in the product $\big(\phi({\bf 78})F({\bf 27})\big)_{\bf 27}$ is the
same as that of the $E_6$ generators. Moreover, the components in 
$\phi({\bf 78})$ possessing the above non-vanishing VEVs, $\chi_R$,
$\chi_L$ and $\omega$, correspond to the Cartan generators 
\begin{equation}
  T_{8 L,R}=\pmatrix{1 & 0 & 0 \cr 0 & 1 & 0 \cr 0 & 0 & -2}
  \in SU(3)_{L,R}, \qquad  
  T_{3R}=\pmatrix{1 & 0 & 0 \cr 0 & -1 & 0\cr 0 & 0 & 0}\in SU(3)_{R}, 
\end{equation}
respectively, so 
that $F({\bf 27}) \to \big(\langle\phi({\bf 78})\rangle 
F({\bf 27})\big)_{\bf 27}$ is equivalent to the replacement of each
component field in $F({\bf 27})$ by the same field multiplied by 
its $T_{8 L,R}$ and $T_{3R}$ quantum numbers and the corresponding
VEVs, $\chi_R$, $\chi_L$, $\omega$. Thus the effective Yukawa
couplings that result from Eq.~(\ref{phi}) via the VEV 
$\langle\phi({\bf 78})\rangle/M_P =\chi_R T_{8R}+\chi_L T_{8L}+\omega 
T_{3R}$ can be found directly from the component expression for the
original direct Yukawa 
coupling $\Psi_i({\bf 27})\Psi_j({\bf 27})H({\bf 27})$ simply by
making this replacement.

The $T_{8 L,R}$ and $T_{3R}$ quantum numbers of the components of 
{\bf 27} can easily be found by considering its decomposition under
the subgroup $SU(3)_L\times SU(3)_R\times SU(3)_C\subset E_6$ (given
explicitly in the Appendix of Ref.~\cite{BK}). We thus
find that the replacement $\Psi({\bf 27})\to M_P^{-1} 
(\langle\phi({\bf 78})\rangle\Psi({\bf 27}))_{\bf 27}$ explicitly
reads as follow for the component fields of $\Psi({\bf 27})$:
\begin{eqnarray}
  \Psi({\bf 3,1,3}) &=& \pmatrix{u_i\cr d_i\cr D_i\cr} \;\to\; 
  \pmatrix{(\chi_L)u_i\cr (\chi_L)d_i\cr (-2\chi_L)D_i\cr}, 
  \label{313}\\[2mm]
  \Psi({\bf 1,3^*,3^*}) &=& \pmatrix{u^{c\,i}\cr d^{c\,i}\cr D^{c\,i}
    \cr} \;\to\; 
  \pmatrix{(-\omega-\chi_R)u^{c\,i}\cr (\omega-\chi_R)d^{c\,i}\cr
    (+2\chi_R)D^{c\,i}\cr}, \\[2mm]
  \Psi({\bf 3^*,3,1}) &=& 
  \bordermatrix{ 
    & 1^*_L & 2^*_L  & 3^*_L \cr
    1_R  &  N^c & E^c  & -e^c \cr
    2_R  &  -E  &  N   & \nu^c \cr
    3_R  &   e  & -\nu&  -S  \cr} \;\to\;
  \bordermatrix{ 
    & -\chi_L & -\chi_L  & +2\chi_L \cr
    \hfil\omega+\chi_R  &  N^c & E^c  & -e^c \cr
    -\omega+\chi_R  &  -E  &  N   & \nu^c \cr
    \hfil-2\chi_R  &   e  & -\nu&  -S  \cr} \nonumber \\[3mm]
  &&\hspace{-2.7cm} =
  \pmatrix{ (\omega+\chi_R-\chi_L)N^c &
    (\omega+\chi_R-\chi_L)E^c & (\omega+\chi_R+2\chi_L)(-e^c) \cr
    (-\omega+\chi_R-\chi_L)(-E) & (-\omega+\chi_R-\chi_L)N &
    (-\omega+\chi_R+2\chi_L)\nu^c \cr (-2\chi_R-\chi_L)\,e  &
    (-2\chi_R-\chi_L)(-\nu) & (-2\chi_R+2\chi_L)(-S) \cr}.
\hspace{3em}
  \label{331}
\end{eqnarray}

In the antisymmetric coupling, $\phi({\bf 78})$ acts on the Higgs
field $H({\bf 27})$. The neutral Higgs components 
$H({\bf 1})$, $H({\bf 16,1})$, $H({\bf 16,5}^*)$, $H({\bf10,5^*})$ and
$H({\bf10,5})$ correspond to the components $S$, $\nu^c$, $-\nu$,
$-N$, and $-N^c$, respectively, of $\Psi({\bf 27})$ and carry the same
quantum numbers as theirs given in Eq.~(\ref{331}). Thus, for
$(\langle\phi({\bf 78})\rangle H({\bf 27}))_{\bf 27}$, we have
only to make the following replacement:
\begin{eqnarray}
  H({\bf 1}) \quad &\to& \quad (-2\chi_R+2\chi_L)H({\bf 1}),
  \nonumber \\
  H({\bf 16,1}) \quad &\to& \quad (-\omega+\chi_R+2\chi_L)
  H({\bf 16,1}),\nonumber \\
  H({\bf 16,5}^*) \quad &\to& \quad (-2\chi_R-\chi_L)H({\bf 16,5}^*),
  \nonumber \\ 
  H({\bf 10,5^*}) \quad &\to& \quad (-\omega+\chi_R-\chi_L)
  H({\bf 10,5^*}), \nonumber \\ 
  H({\bf 10,5}) \quad &\to& \quad (\omega+\chi_R-\chi_L)H({\bf 10,5}).
  \label{replace}
\end{eqnarray}

The direct Yukawa coupling of the $y_{ij}$ term in Eq.~(\ref{yukawa}) 
is given by
\begin{eqnarray}
  W_Y(H) &=& y_{ij}\left[ -H({\bf 10,5})\left(u^c_i u_j +\nu^c_i \nu_j 
      -S_i N_j \right) -H({\bf 10,5}^*) \left( d_i d^{c}_j +e^c_i e_j
    \right) \right. \nonumber \\[1mm]
  &&\left.\hspace{.8cm} +H({\bf 16,5}^*) \left(d_i D^{c}_j +e^c_i E_j
      \right) -H({\bf 16,1}) \left(D_i d^c_j +E^c_i e_j \right)\right].
    \label{WY}
\end{eqnarray}
Now making in this expression the above replacements,
(\ref{313})--(\ref{331}) and (\ref{replace}), we obtain the following
explicit expression for the effective Yukawa 
coupling $W_{\langle\phi\rangle}$ resulting from Eq.~(\ref{phi}):
\begin{eqnarray}
  W_{\langle\phi\rangle}
  &=& -H({\bf 10,5}) \biggl[\Bigl(-\frac{1}{2} 
  (\chi_R+\chi_L+\omega) a_{ij} +(\omega+\chi_R-\chi_L)s_{ij}\Bigr)
  u^c_i u_j \nonumber \\ 
  &&\hspace{6em} +\Bigl(\frac{1}{2}
  (3\chi_R+3\chi_L-\omega)a_{ij} +(\omega+\chi_R-\chi_L)s_{ij} 
  \Bigr)\nu^c_i \nu_j \nonumber \\
  &&\hspace{6em} +\Bigl(\frac{1}{2}(\omega-3\chi_R+3\chi_L)a_{ij} 
  +(\omega+\chi_R-\chi_L)s_{ij} \Bigr)S_i N_j \biggr] \nonumber \\
  && -H({\bf 10,5}^*) \biggl[\Bigl(\frac{1}{2}
  (\chi_R+\chi_L-\omega)a_{ij}+(\chi_R-\chi_L-\omega)s_{ij}\Bigr)
  d_i d^{c}_j \nonumber \\
  &&\hspace{6em} +\Bigl(\frac{1}{2}(3\chi_R+3\chi_L+\omega)a_{ij} 
  +(\chi_R-\chi_L-\omega)s_{ij}\Bigr)e^c_i e_j \biggr] \nonumber \\
  && +H({\bf 16,5}^*) \biggl[\Bigl(-\frac{1}{2}(2\chi_R-\chi_L)a_{ij}
  -(2\chi_R+\chi_L)s_{ij}\Bigr)d_i D^{c}_j \nonumber \\
  &&\hspace{6em} +\Bigl((\omega+{3\over2}\chi_L)a_{ij}
  -(2\chi_R+\chi_L) s_{ij}\Bigr) e^c_i E_j \biggr] \nonumber \\
  && -H({\bf 16,1}) \biggl[\Bigl(\frac{1}{2}(\chi_R-2\chi_L-\omega)
  a_{ij} +(-\omega+\chi_R+2\chi_L)s_{ij}\Bigr)D_i d^{c}_j \nonumber\\ 
  &&\hspace{6em} +\Bigl(\frac{1}{2}(\omega+3\chi_R)a_{ij}
  +(-\omega+\chi_R+2\chi_L) s_{ij}\Bigr) E^c_i e_j \biggr].
  \label{Wphi}
\end{eqnarray}
In the expressions (\ref{WY}) and (\ref{Wphi}), and also henceforth in
this paper, the coupling constants $y_{ij}$, $s_{ij}$ and $a_{ij}$ are
no longer the original ones (which were all of order 1), but should be
understood as representing the following coupling constants suppressed 
by the powers of $\lambda=\langle\Theta\rangle/M_P$: 
\begin{equation}
  \Bigl(g_{ij}\Bigr) = \Bigl(g^{\rm org}_{ij}\lambda^{f_i+f_j}\Bigr) = 
  \pmatrix{g^{\rm org}_{11}\lambda^6 & g^{\rm org}_{12}\lambda^5 &
  g^{\rm org}_{13}\lambda^3 \cr g^{\rm org}_{21}\lambda^5 & g^{\rm
  org}_{22}\lambda^4 & g^{\rm org}_{23}\lambda^2 \cr g^{\rm
  org}_{31}\lambda^3 & g^{\rm org}_{32}\lambda^2 & g^{\rm org}_{33}
  \cr} \quad {\rm for\ }\ g=y,\ s,\ a.
\end{equation}
It is therefore important to realize that the coupling 
constants $y_{ij}$, $s_{ij}$ and $a_{ij}$, as well as their linear
combinations, like $Y_{ij}$ and $Y'_{ij}$ defined below, are
quantities of order $\lambda^{f_i+f_j}$ ($f_1=3,\ f_2=2,\ f_3=0$).

\section{Effective Yukawa couplings}
\setcounter{equation}{0}

As we mentioned above, since the $SU(5)$ ${\bf 10}$ component
appears only once in $\Psi({\bf 27})$, the identification of the three
up-type quarks is unique. Thus the effective Yukawa texture for
up-type quarks, induced after the VEVs of $\Theta$ 
and $\phi({\bf 78})$ are developed, can immediately be written down in
terms of $y_{ij}$, $s_{ij}$ and $a_{ij}$ as
\begin{equation}
  \bordermatrix{
    & u_1^c & u_2^c & u_3^c \cr
    u_1 & Y_{11} & Y_{12} & Y_{13} \cr
    u_2 & Y_{21} & Y_{22} & Y_{23} \cr
    u_3 & Y_{31} & Y_{32} & y_{33} \cr},
\end{equation}
where 
\begin{equation}
  Y_{ij}=y_{ij}+(\chi_R-\chi_L+\omega)s_{ij}
  +\frac{1}{2}(\chi_R+\chi_L+\omega)a_{ij}.
\end{equation}
(The quantity $a_{ii}$ is 0 by definition.) \ Note that since we have
assumed that the $U(1)$ charge of the $\phi({\bf 78})$ Higgs field is
$-2$, the additional Yukawa couplings, $s_{ij}$ and $a_{ij}$, coming
from $\phi({\bf 78})$ do not contribute to the 3-3 element. Note also
that this texture is not necessarily symmetric, because of the
asymmetric coupling from $\phi({\bf 78})$.

\begin{figure}[htbp]
  \epsfxsize= 7cm
  \centerline{\epsfbox{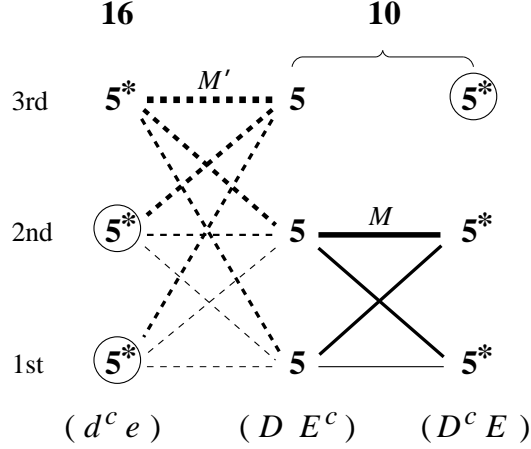}}
  \caption{The E-twisted structure in which the ${\bf 5}^*$ components
    enclosed by circles are the dominant components that remain
    light. The solid and dotted lines indicate heavy mass terms given
    by $\langle\Phi({\bf 1})\rangle$ and 
    $\langle H({\bf 16,1})\rangle$, respectively. The
    $U(1)$-suppressed terms with higher powers of 
    $\langle\Theta\rangle/M_P$ are indicated by the thinner lines.}
  \label{decouple}
\end{figure}

As for the down-quark and charged-lepton sectors, the situation is
more complicated, since we now utilize the E-twisted structures
(\ref{twist}). As explained in Ref.~\cite{BK}, the E-twisted structure
is realized when $\Phi({\bf 1})$ and $H({\bf 16,1})$ develop the
following VEVs:
\begin{equation}
  \langle\Phi({\bf 1})\rangle\simeq M, \qquad
  \langle H({\bf 16,1})\rangle\simeq M'.
\end{equation}
$\langle\Phi({\bf 1})\rangle$ gives mass terms 
for $\Psi_i({\bf 10,5^*})$-$\Psi_j({\bf 10,5})$ and 
$\langle H({\bf 16,1})\rangle$ does so 
for $\Psi_i({\bf 16,5^*})$-$\Psi_j({\bf 10,5})$. Noting that $\Phi$
and $H$ have $U(1)$ charges $-4$ and 0, respectively, the superheavy
mass terms are formed in the pattern depicted in Fig.~\ref{decouple}.
Now assume that $M\gg M'$, implying that the breaking scale 
of $E_6$ down to $SO(10)$ is higher than the breaking scale 
of $SO(10)$ to $SU(5)$. With this natural assumption, the components 
$\Psi_{1,2}({\bf 10,5^*})=(D^c_{1,2}, E_{1,2})$ decouple from the light
sector by forming superheavy mass terms 
with $\Psi_{1,2}({\bf 10,5})=(D_{1,2}, E^c_{1,2})$ of order $M$. Thus
$D^c_3$ and $E_3$ in $\Psi_3({\bf 10,5^*})$ remain very light
(massless at this stage), and we can identify them as the third
generation down quark and charged lepton. The first and second
generations of ${\bf 5}^*$, on the other hand, are not 
exactly $\Psi_{1,2}({\bf 16,5^*})=(d^c_{1,2}, e_{1,2})$, but are
slightly mixed with the third generational one, 
$\Psi_3({\bf 16,5^*})$, owing to the mass mixing with 
$\Psi_3({\bf 10,5})=(D_3, E^c_3)$. $\Psi_3({\bf 10,5})=(D_3, E^c_3)$
forms the following mass terms by taking account also of the
contributions from the $\phi({\bf 78})$ Higgs:
\begin{eqnarray}
  &&\hspace*{-5mm} 
  M'D_3\Bigl((Y_{31}-2\omega\st{31}+3\chi_L\st{13})d^c_1 
  +(Y_{32}-2\omega\st{32}+3\chi_L\st{23})d^c_2+ y_{33}d^c_3\Bigr)
  \nonumber \\[1mm]
  &&\hspace*{-5mm} 
  {}+M'E^c_3\Bigl((Y'_{31}-2\omega\st{31}+3\chi_L\st{13})e_1 
  +(Y'_{32}-2\omega\st{32}+3\chi_L\st{23})e_2+y_{33}e_3\Bigr),
  \label{D3mass}
\end{eqnarray}
where $\st{ij}\equiv s_{ij}+\frac{1}{2}a_{ij}$ and 
\begin{eqnarray}
  Y'_{ij} &\equiv& y_{ij}+(\chi_R-\chi_L+\omega)s_{ij}
  +\frac{3}{2}(\chi_R+\chi_L+\omega)a_{ij} \\[1mm]
  &=& Y_{ij} +(\chi_R+\chi_L+\omega)a_{ij}. \nonumber
\end{eqnarray}
Thus the superheavy mass partners of $D_3$ and $E^c_3$ are not 
exactly $d^c_3$ and $e_3$ but the linear combinations appearing in
Eq.~(\ref{D3mass}). The light down quarks and charged leptons for 
the first two generations should be orthogonal to these and are,
therefore, given by
\begin{eqnarray}
  d'^c_1 &=& d^c_1 -\frac{1}{y_{33}}
  (Y_{31}-2\omega\st{31}+3\chi_L\st{13}) d^c_3,\nonumber \\
  d'^c_2 &=& d^c_2 -\frac{1}{y_{33}}
  (Y_{32}-2\omega\st{32}+3\chi_L\st{23}) d^c_3, \label{dmix} \\
  e'_1 &=& e_1 -\frac{1}{y_{33}}
  (Y'_{31}-2\omega\st{31}+3\chi_L\st{13}) e_3,\nonumber \\
  e'_2 &=& e_2 -\frac{1}{y_{33}}
  (Y'_{32}-2\omega\st{32}+3\chi_L\st{23}) e_3,
\end{eqnarray}
up to unimportant minor mixings with heavier components. Using these
sets of the light fields of three generations, we can obtain the final
forms of texture for the down-quark and charged-lepton Yukawa
couplings.

The down quarks $d_i$-$d^c_j$ have the Yukawa couplings
\begin{equation}
  \bordermatrix{
    & d^c_1 & d^c_2 & d^c_3 \cr
    d_1 & (Y_{11}-2\omega\st{11}) & (Y_{12}-2\omega\st{12}) &
    (Y_{13}-2\omega\st{13}) \cr  
    d_2 & (Y_{21}-2\omega\st{21}) & (Y_{22}-2\omega\st{22}) &
    (Y_{23}-2\omega\st{23}) \cr 
    d_3 & (Y_{31}-2\omega\st{31}) & (Y_{32}-2\omega\st{32}) & y_{33}
    \cr}
\end{equation}
from $H({\bf 10,5^*})$, and $d_i$-$D^c_j$ have
\begin{equation}
  \bordermatrix{
    & D^c_1 & D^c_2 & D^c_3 \cr
    d_1 & Y_{11}-(\omega+3\chi_R)\st{11} &
    Y_{12}-(\omega+3\chi_R)\st{12} & 
    Y_{13}-(\omega+3\chi_R)\st{13} \cr  
    d_2 & Y_{21}-(\omega+3\chi_R)\st{21} &
    Y_{22}-(\omega+3\chi_R)\st{22} & 
    Y_{23}-(\omega+3\chi_R)\st{23} \cr 
    d_3 & Y_{31}-(\omega+3\chi_R)\st{31} &
    Y_{32}-(\omega+3\chi_R)\st{32} & y_{33} \cr}
\end{equation}
from $H({\bf 16,5^*})$. Now, taking account of the mixing (\ref{dmix})
between the $d^c_i$ and the fact that the light Higgs doublet 
$H({\bf 5}^*)$ in the low-energy region is given by the linear
combination $H({\bf 10,5^*})\cos\theta+H({\bf 16,5^*})\sin\theta$, the
Yukawa matrix for the light down quarks is found to be
\begin{equation}
  \bordermatrix{
    & d'^c_1 & d'^c_2 & D^c_3 \cr
    d_1 & Y^d_{11}\cos\theta& Y^d_{12}\cos\theta&
    (Y_{13}-(\omega+3\chi_R)\st{13}) \sin\theta\cr
    d_2 & Y^d_{21}\cos\theta& Y^d_{22}\cos\theta&
    (Y_{23}-(\omega+3\chi_R)\st{23}) \sin\theta\cr
    d_3 & -3\chi_L\st{13}\cos\theta& -3\chi_L\st{23}\cos\theta&
    y_{33}\sin\theta\cr}, 
\end{equation}
where
\begin{equation}
  Y^d_{ij} \equiv Y_{ij}-2\omega\st{ij} -\frac{1}{y_{33}}
  (Y_{i3}-2\omega\st{i3})(Y_{3j}-2\omega\st{3j}+3\chi_L\st{j3}).
\end{equation}
This is the final form of the down-quark Yukawa coupling in our
model. The charged-lepton matrix is obtained in the same way. However,
comparing all the Yukawa couplings for the down-quark and
charged-lepton sectors, we obtain the rule that all the charged-lepton
couplings can be derived by making the replacement $Y_{ij}\to Y'_{ij}$
from the corresponding down-quark couplings. Applying this rule, we
immediately obtain the final form of Yukawa coupling for the three
light generations of charged-leptons:
\begin{equation}
  \bordermatrix{
    & e'_1 & e'_2 & E_3 \cr
    e^c_1 & Y^e_{11}\cos\theta& Y^e_{12}\cos\theta&
    (Y'_{13}-(\omega+3\chi_R)\st{13}) \sin\theta\cr
    e^c_2 & Y^e_{21}\cos\theta& Y^e_{22}\cos\theta&
    (Y'_{23}-(\omega+3\chi_R)\st{23}) \sin\theta\cr
    e^c_3 & -3\chi_L\st{13}\cos\theta& -3\chi_L\st{23}\cos\theta&
    y_{33}\sin\theta\cr}, 
\end{equation}
where $Y^e$ is defined as
\begin{equation}
  Y^e_{ij} \equiv Y'_{ij}-2\omega\st{ij} -\frac{1}{y_{33}}
  (Y'_{i3}-2\omega\st{i3})(Y'_{3j}-2\omega\st{3j}+3\chi_L\st{j3}).
\end{equation}

\section{Mass eigenvalues and mixings}
\setcounter{equation}{0}

In this section, using the above sets of Yukawa couplings for the
three light generations, we derive the total $3\times3$ mass textures
for the up-quark, down-quark and charged-lepton, $M_u$, $M_d$ 
and $M_e$, and obtain the mass eigenvalues and mixing angles. First we
summarize the mass matrices obtained in the previous section:

\noindent
{\it mass matrix for $u$}
\begin{equation}
  M_u = \bordermatrix{
    & u^c_1 & u^c_2 & u^c_3 \cr
    u_1 & Y_{11} & Y_{12} & Y_{13} \cr
    u_2 & Y_{21} & Y_{22} & Y_{23} \cr
    u_3 & Y_{31} & Y_{32} & 1 \cr}
  y_{33}v \sin\beta, 
  \label{mu}
\end{equation}
{\it mass matrix for $d$}
\begin{equation}
  M_d = \bordermatrix{
    & d'^c_1 & d'^c_2 & D^c_3 \cr
    d_1 & Y^d_{11}\cos\theta& Y^d_{12}\cos\theta&
    (Y_{13}-(\omega+3\chi_R)\st{13}) \sin\theta\cr
    d_2 & Y^d_{21}\cos\theta& Y^d_{22}\cos\theta&
    (Y_{23}-(\omega+3\chi_R)\st{23}) \sin\theta\cr
    d_3 & -3\chi_L\st{13}\cos\theta& -3\chi_L\st{23}\cos\theta&
    \sin\theta\cr} 
  y_{33}v\cos\beta,
  \label{md}
\end{equation}
{\it mass matrix for $e$}
\begin{equation}
  M_e^{\rm T} =
  \bordermatrix{
    & e'_1 & e'_2 & E_3 \cr
    e^c_1 & Y^e_{11}\cos\theta& Y^e_{12}\cos\theta&
    (Y'_{13}-(\omega+3\chi_R)\st{13}) \sin\theta\cr
    e^c_2 & Y^e_{21}\cos\theta& Y^e_{22}\cos\theta&
    (Y'_{23}-(\omega+3\chi_R)\st{23}) \sin\theta\cr
    e^c_3 & -3\chi_L\st{13}\cos\theta& -3\chi_L\st{23}\cos\theta&
    \sin\theta\cr} 
  y_{33}v\cos\beta. 
  \label{me}
\end{equation}
In the above, $\tan\beta$ is the mixing angle of two light Higgs
doublets and $v$ is the vacuum expectation value of the standard model
Higgs field. Here we have redefined all the Yukawa couplings to be
normalized by $y_{33}$, but we have used the same notation $Y_{ij}$ as
above, in order to avoid an overabundance of parameters. As a
convention for the mass matrix $M$, we have assumed a fermion mass
term given in the form 
${\cal L}_{\rm mass}=\bar\psi_{{\rm L}\,i}M_{ij}
\psi_{{\rm R}\,j}+{\rm h.c.}$, which explains why we have applied the
transpose T to the charged-lepton matrix $M_e$. 

These mass matrices are diagonalized as 
\begin{eqnarray}
  (M_u)_{\rm diag} = U_u M_u V_u^\dagger,\qquad 
  (M_d)_{\rm diag} = U_d M_d V_d^\dagger,\qquad 
  (M_e)_{\rm diag} = U_e M_e V_e^\dagger,
\end{eqnarray}
and then the Cabibbo-Kobayashi-Maskawa (CKM) and Maki-Nakagawa-Sakata
(MNS) mixing matrices are given by
\begin{eqnarray}
  V_{\rm CKM} = U_u U_d^\dagger,\qquad 
  V_{\rm MNS} = U_e U_\nu^\dagger,
\end{eqnarray}
where $U_\nu$ is such that it makes the matrix 
$U_\nu^* M_\nu U_\nu^\dagger$ diagonal. The matrix $M_\nu$ is the
Majorana mass matrix of the light left-handed neutrinos, which we do
not explicitly discuss in this paper. However, we here only note that
it typically takes a hierarchical form~\cite{BK} like
\begin{equation}
  M_\nu\propto 
  \pmatrix{
    \lambda^{4} & \lambda^{3} & \lambda^{2} \cr
    \lambda^{3} & \lambda^{2} & \lambda^{1} \cr
    \lambda^{2} & \lambda^{1} &  1     \cr}\,.
\end{equation}
Thus the mixing matrix on the neutrino side is almost 
unity, $U_\nu\sim 1 + O(\lambda)$, and hence the mixing matrix in the
charged-lepton side $U_e$ is equal to the MNS matrix $V_{\rm MNS}$, up
to $O(\lambda)$ corrections.

It is straightforward to obtain the mass eigenvalues and mixing matrix
for each mass matrix. However, we should take some care in treating
the matrices for the down quarks and charged leptons, since there
occurs a cancellation of the terms of leading order in $\lambda$. We
present in the Appendix explicit formulas for eigenvalues and mixing
matrices in such a case. Applying the formulas there to
Eqs.~(\ref{mu})--(\ref{me}), we find the mass eigenvalues
\begin{eqnarray}
  m_t &=& y_{33}\,v \sin\beta, \nonumber \\
  m_c &=& y_{33}\tu{22}\,v \sin\beta, \nonumber \\
  m_u &=& y_{33}\left(\tu{11}-\frac{\tu{12}\tu{21}}{\tu{22}}\right)
  v\sin\beta, \nonumber \\
  m_b &=& y_{33}\, S\sin\theta\, v\cos\beta, \nonumber \\
  m_s &=& y_{33}\frac{\td{22}}{S}\cos\theta\, v \cos\beta,\nonumber\\
  m_d &=& y_{33}\left(\td{11}-\frac{\td{12}\td{21}}{\td{22}}\right)
  \cos\theta\, v \cos\beta, \nonumber \\
  m_\tau&=& y_{33}\, S\sin\theta\, v \cos\beta,\;\; (=m_b) \nonumber\\
  m_\mu&=& y_{33}\frac{\te{22}}{S} \cos\theta\, v \cos\beta,
  \nonumber \\ 
  m_e &=& y_{33}\left(\te{11}-\frac{\te{12}\te{21}}{\te{22}}\right) 
  \cos\theta\, v \cos\beta,
\end{eqnarray}
and the CKM and MNS matrix elements
\begin{eqnarray}
  V_{us} &=& \frac{\td{12}}{\td{22}}-\frac{\tu{12}}{\tu{22}},
  \nonumber \\
  V_{ub} &=& \frac{-1}{S^2\tu{22}}\left[(3\chi_R+\omega+
    18\omega\chi_L^2\st{23}^2 \cot^2\theta) \left(\st{13}\tu{22}
      -\st{23}\tu{12}\right) \right. \nonumber\\
  &&\hspace{2cm}\left. +6\,\omega\chi_L\st{23}\cot^2\theta 
    \left(f_{12}\tu{22}-f_{22}\tu{12}\right)\right], \nonumber \\
  V_{cb} &=& \frac{-\st{23}}{S^2}\left[3\chi_R+w+3\chi_L\cot^2\theta 
    \left(\td{22}+3\chi_L(3\chi_R+\omega)\st{23}^2\right)\right],
  \nonumber \\
  V_{e2} &=& -\frac{\te{21}}{\te{22}}+(U^\dagger_\nu)_{12},\nonumber\\
  V_{e3} &=& \frac{3\chi_L\cot\theta}{\te{22}}\left(\st{13}\te{22}
    -\st{23}{\te{21}}\right), \nonumber \\
  V_{\mu3} &=& \frac{3\chi_L\st{23}\cot\theta}{S}.
\end{eqnarray}
Here $S$, $T_{ij}^{u,d,e}$ and $f_{ij}$ are given by the following
combinations of the coupling constants:
\begin{eqnarray}
  \tu{ij} &\equiv& Y_{ij} -Y_{i3}Y_{3j},\\
  \td{ij} &\equiv& Y_{ij}-2\omega\st{ij}
  -(Y_{i3}-2\omega\st{i3})(Y_{3j}-2\omega\st{3j})
  -3\chi_L(3\chi_R-\omega)\st{i3}\st{j3}, \\
  \te{ij} &\equiv& Y'_{ij}-2\omega\st{ij}
  -(Y'_{i3}-2\omega\st{i3})(Y'_{3j}-2\omega\st{3j})
  -3\chi_L(3\chi_R-\omega)\st{i3}\st{j3}, \\
  S &\equiv& \left(1+9\chi_L^2\st{23}^2\cot^2\theta\right)^{1/2}, \\
  f_{ij} &\equiv& \st{ij} -\st{i3}(Y_{3j}-\omega\st{3j})
  -\st{3j}(Y_{i3}-\omega\st{i3}).
\end{eqnarray}
Note that in these equations, all the Yukawa couplings are normalized
by $y_{33}$, and also that $T_{ij}^{u,d,e}$ and $f_{ij}$ are
$O(\lambda^{f_i+f_j})$ quantities. It is interesting that these
eigenvalues and mixing angles depend on the Yukawa couplings only
through their particular combinations, like $T_{ij}^{u,d,e,}$ and
$f_{ij}$. Therefore, although there are many independent Yukawa
couplings in the present model, the actual number of free parameters
is greatly reduced. Indeed, as seen in the next section, the number of 
parameters is less than the number of experimentally measured
quantities, and hence we can obtain a kind of sum rule for the
observables. The lepton 1-2 mixing $V_{e2}$ cannot be predicted,
because in the present model, the 1-2 mixing from the neutrino side is 
comparable to that from the charged-lepton side, and we are not
specifying the contributions of higher-dimensional operators in the
neutrino sector. In the following, we examine whether the low-energy
experimental values can be properly reproduced with this small number
of free parameters.

\section{Predictions}
\setcounter{equation}{0}

We assume in this paper that below the GUT scale, the model is
described by the minimal supersymmetric standard model down to the
low-energy supersymmetry breaking scale. In order to compare the
predictions of our model with the experimental data, it is convenient
to analyze the mass matrices at the GUT scale. We here present the
numerical values of the running quark and lepton masses and mixing
angles at a scale $\simeq 2\times10^{16}$ GeV~\cite{exp}:
\begin{equation}
  \begin{array}{lll}
    m_u \,\sim\, 1.04_{\;-0.20}^{\;+0.19} {\rm ~MeV}, &
    m_d \,\sim\, 1.33_{\;-0.19}^{\;+0.17} {\rm ~MeV}, &
    m_e \,\sim\, 0.325 {\rm ~MeV}, \\
    m_c \,\sim\, 302_{\;-27}^{\;+25} {\rm ~MeV}, &
    m_s \,\sim\, 26.5_{\;-3.7}^{\;+3.3} {\rm ~MeV}, &
    m_\mu\,\sim\, 68.6 {\rm ~MeV}, \\
    m_t \,\sim\, 129_{\;-40}^{+196} {\rm ~GeV}, &
    m_b \,\sim\, 1.00\pm0.04 {\rm ~GeV}, &
    m_\tau\,\sim\, 1.17 {\rm ~GeV},
  \end{array}
\end{equation}
\begin{equation}
  V_{us} = 0.217-0.224,\qquad V_{cb} = 0.031-0.037,\qquad
  V_{ub} = 0.002-0.005.
\end{equation}

\subsection{Second and third generations}
\begin{figure}[htbp]
  \begin{center}
    \leavevmode
    \epsfxsize=11cm \ \epsfbox{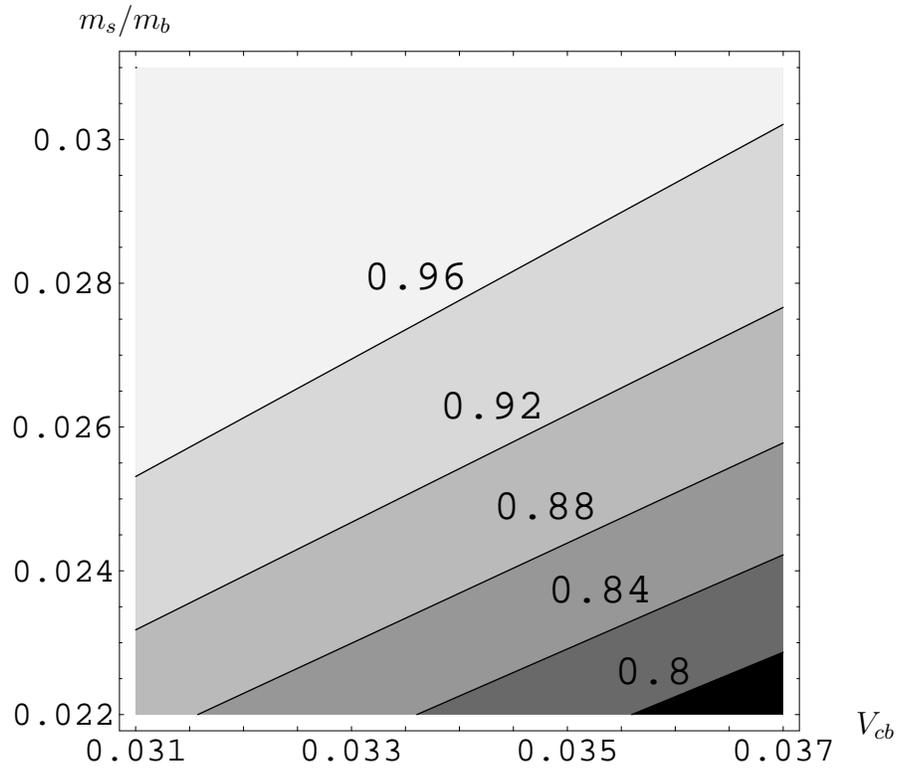}
    \put(-286,285){$m_s/m_b$}
    \put(8,18){$V_{cb}$}
    \caption{The prediction of the lepton 2--3 mixing angle 
      $\sin^2 2\theta_{\mu\tau}$ from our relation
      (\ref{sumrule}). This square parameter range represents the
      experimental uncertainties of $m_s/m_b$ and $V_{cb}$. In almost
      the entire region, the relation is consistent with the
      observations.}
    \label{sum}
  \end{center}
\end{figure}

First let us concentrate on the second and third generations. As is
obvious from Eqs.~(\ref{md}) and (\ref{me}), the 3-1 and 3-2 matrix
elements of $M_d$ and $M_e^{\rm T}$ vanish unless $\chi_L\neq 0$, and
therefore even for the simplest option we need a nonzero value of
$\chi_L$. We first consider the simplest case $\chi_R=\omega=0$. The
number of free parameters is then greatly reduced. The mass
eigenvalues and mixing angles can be written in the following simple
forms:
\begin{eqnarray}
  m_t &=& y_{33}\, v \sin\beta, \nonumber \\
  m_c &=& y_{33}T_{22}\, v \sin\beta, \nonumber \\
  m_b &=& y_{33}S v\sin\theta\cos\beta, \nonumber \\
  m_s &=& y_{33}\frac{T_{22}}{S} v\cos\theta\cos\beta, \nonumber \\
  m_\tau&=& y_{33}S v\sin\theta\cos\beta, \nonumber \\
  m_\mu&=& y_{33}\frac{T_{22}+2(\chi_L a_{23})^2}{S} v\cos\theta 
  \cos\beta,
  \label{23mass}
\end{eqnarray}
\begin{eqnarray}
  V_{cb} &=& \frac{-3\chi_L\st{23}T_{22}}{S^2} \cot^2 \theta, \\
  V_{\mu3} &=& \frac{3\chi_L\st{23}\cot\theta}{S}.
  \label{23mix}
\end{eqnarray}
Here we have denoted $\tu{22}=\td{22}$ by $T_{22}$. In this case,
there are 7 independent free parameters in all in the second and third
generation part, $y_{33}$, $T_{22}$, $\st{23}$, $a_{23}$, $\theta$,
$\tan\beta$ and $\chi_L$. However, since the VEV $\chi_L$ always
appears multiplied by the Yukawa coupling $\st{23}$ or $a_{23}$, the
number of parameters is in essence 6, with which we express all the
relevant data of masses and mixings. We can `solve' the above
relations (\ref{23mass})--(\ref{23mix}) and inversely express the
parameters in terms of the observable quantities:
\begin{eqnarray}
  y_{33} &=& \frac{m_t}{v}\frac{1}{\sin\beta}, \\[1mm]
  T_{22} &=& \frac{m_c}{m_t}, \label{T}\\
  \st{23}\chi_L &=& -\frac{1}{3}\cdot
  \frac{\displaystyle\frac{m_c}{m_t}V_{cb}}{{V_{cb}^2
      +\left(\displaystyle\frac{m_s}{m_b}\right)^2}}, \\
  (a_{23}\chi_L)^2 &=& \frac{1}{2}\frac{m_c}{m_t} 
  \left(\frac{m_\mu}{m_s}-1\right), \\[1mm]
  \tan\theta&=& \frac{\displaystyle\frac{m_c}{m_t}
    \frac{m_s}{m_b}}{V_{cb}^2 
    +\left(\displaystyle\frac{m_s}{m_b}\right)^2}, \\
  \tan\beta&=& \frac{m_t}{m_s}\sqrt{V_{cb}^2
  +\left(\displaystyle\frac{m_s}{m_b}\right)^2} \sin\theta.
\end{eqnarray}
Now there are 8 observables, $m_t$, $m_c$, $m_b$, $m_s$, $m_\tau$,
$m_\mu$, $V_{cb}$ and $V_{\mu3}$, and so there exist two relations
between the observable quantities, which give our
parameter-independent predictions. One of them is not surprising:
$m_b=m_\tau$. This is the well-known $SU(5)$ GUT
relation~\cite{btau}. The other is a novel relation connecting the
quark and lepton mixing angles,
\begin{eqnarray}
  \sin^2 2\theta_{\mu\tau} &=& \frac{4|V_{cb}|^2
    \left(\displaystyle\frac{m_s}{m_b}\right)^2}{\left[|V_{cb}|^2   
      +\left(\displaystyle\frac{m_s}{m_b}\right)^2\right]^2},
  \label{sumrule}
\end{eqnarray}
where we have used $\tan \theta_{\mu\tau}=V_{\mu 3}/V_{\tau 3}$. It
should be kept in mind that precisely speaking the left-hand 
side $\theta_{\mu\tau}$ is the charged-lepton mixing angle, which may
deviate from the MNS mixing angle by a neutrino-side contribution of
$\lsim O(\lambda)$. Also the masses $m_b$ and $m_s$ and the CKM
element $V_{cb}$ on the right-hand side are the quantities now
discussed up to $O(\lambda^2)$ corrections. Aside from these
uncertainties, we can predict the $\nu_\mu$--$\nu_\tau$ mixing angle
from only the quark part information. In Fig.~\ref{sum}, we give a
comparison of the relation with the experimental data.
{}From the figure, we can see that the relation (\ref{sumrule}) is
quite consistent with the experimental values.

Note that the success of this relation is a characteristic feature of
the twisting ${\bf5}^*$ structure in this model. To see this, let us
consider the mass matrices
\begin{eqnarray}
  M_u \,\propto\, \bordermatrix{
    & 10_2 & 10_3 \cr
    10_2 & & x \cr
    10_3 & & 1 \cr}, \quad
  M_d \,\propto\, \bordermatrix{
    & 5^*_2 & 5^*_3 \cr
    10_2 & z & x' \cr
    10_3 & y & 1 \cr}, \quad
  M_e^{\rm T} \,\propto\, \bordermatrix{
    & 5^*_2 & 5^*_3 \cr
    10_2 & & \cr
    10_3 & y & 1 \cr}.
\end{eqnarray}
We have assumed the hierarchical form of the up-type Yukawa
matrix. The element $y$ is common to $M_d$ and $M_e$, due to 
the $SU(5)$ GUT symmetry, and $z$ is related to the $m_s$ mass. The
blank entries are irrelevant to the discussion here. The relation
(\ref{sumrule}) itself results if the condition (i) $x=x'$
holds. However, in order for it to predict a large lepton mixing 
angle $\theta_{\mu\tau}$ (or equivalently, `large' right-hand side of
(\ref{sumrule})), we need the additional condition (ii) 
$y\sim O(1)$. The first condition is satisfied in the present model
due to the fact that ${\bf 10}_3$ and ${\bf 5}^*_3$ come from a common
single multiplet. This implies that we need a unification group
$SO(10)$ or larger. The second condition (ii) is satisfied thanks to
the ${\bf 5}^*$ twisting in the present model. As seen in the Appendix,
the lepton mixing angle is determined solely by the parameter $y$. In
the down-quark matrix side also, an $O(1)$ $y$ affects $V_{cb}$ as
well as $m_s/m_b$. This successful relation is a very interesting and
common feature of the generation (${\bf 5}^*$) twisting mechanism, and 
it is valid also in the $SO(10)$ model considered by Nomura and
Yanagida~\cite{NY}. We would like to emphasize, however, that in the
present $E_6$ model, $y\sim O(1)$ also explains the top-bottom
hierarchy. This is due to the fact that we have twisted not the second
generation, ${\bf 5}^*_2$, but the third one, ${\bf 5}^*_3$.

We also have the inequality
\begin{eqnarray}
  \frac{m_\mu}{m_s} &=& 1+\frac{2(a_{23}\chi_L)^2}{T_{22}} \;\;>1.
\end{eqnarray}
(The sign of $T_{22}$ is positive since it gives $m_c/m_t$ by the
relation Eq.~(\ref{T}).) This relation indicates that $m_\mu$ is
always larger than $m_s$ around the GUT scale. This is indeed required
to reproduce their correct low-energy mass eigenvalues when we take
into account the $SU(3)$ gauge contributions to $m_s$ running in the
renormalization-group evolution down to the electroweak scale. The
enhancement of the $m_\mu$ mass at the GUT scale is known and is built
into the Georgi-Jarlskog-type texture~\cite{GY}.

\begin{figure}[htbp]
  \begin{center}
    \leavevmode
    \epsfxsize=10cm \ \epsfbox{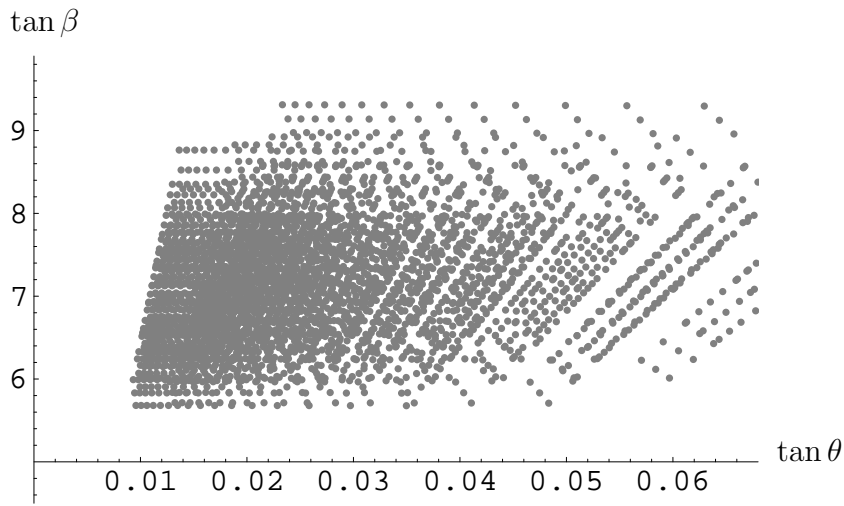}
    \put(7,24){$\tan\theta$}
    \put(-283,187){$\tan\beta$}
    \caption{The allowed regions of the two Higgs mixing angles
      $\tan\theta$ and $\tan\beta$. } 
    \label{tt}
  \end{center}
\end{figure}

To see the dependence on the other parameters, we display the two
Higgs mixing angles $(\tan\theta,\tan\beta)$ plot in
Fig.~\ref{tt}. Each dot satisfies the experimentally observed mass
eigenvalues and mixing angles of the quarks and leptons.
{}From this, we see that the Higgs mixing angle $\theta$ is of 
order $\lambda^2$, as expected from the large lepton mixing. Note that
the model also predicts a small value of $\tan\beta\sim O(1)$, because 
the bottom Yukawa coupling is accompanied by $\sin\theta$. This may be
a beautiful explanation of the reason that $m_b$ is $\lambda^2$ times
smaller than $m_t$. We also find that the other parameters are
naturally consistent with the low-energy experimental values at this
stage.

\subsection{Including the first generation}

In this section, we include the first generation effects in the two
generation analysis of the previous section. Generally speaking,
making analyses of such tiny parameters as the masses of the first
generation might only result in unnecessary details, because, for
example, other (unknown) higher-dimensional operators might become
relevant. Moreover, for the lepton mixing $V_{e3}$, a naive order of
magnitude estimation predicts an order $\lambda^1$ value which is near
the experimental upper limit, according to the CHOOZ
data~\cite{chooz}. Nevertheless, it would be interesting to do this
analysis in the present model defined by the interaction terms
(2.1)--(2.3) alone.

\begin{figure}[htbp]
  \begin{center}
    \leavevmode
    \epsfxsize=10cm \ \epsfbox{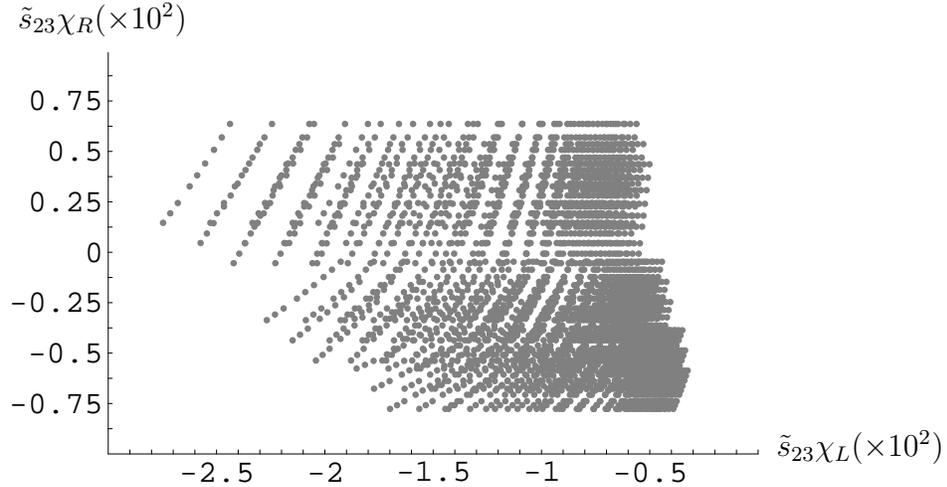}
    \put(7,23){$\st{23}\chi_L (\times10^2)$}
    \put(-280,185){$\st{23}\chi_R (\times10^2)$}
    \caption{Typical allowed region for the adjoint Higgs
      VEVs. $\chi_R$ generally takes a smaller value than
      $\chi_L$. The central experimental values correspond to
      $\st{23}\chi_L\simeq -1.3\times10^{-2}$ 
      and $\st{23}\chi_R\simeq -0.24\times 10^{-2}$.}
    \label{lr}
  \end{center}
\end{figure}

In order to make a full three generation analysis, we have to turn on
the parameters other than $\chi_L$, because the nonzero contribution
from $\chi_L$ alone cannot create any difference between the first
generation mixings of the up and down quarks (see section 5). Thus we
are lead to the next minimum choice to include a nonzero $\chi_R$, 
while keeping $\omega$ set to zero. In this case, $\tu{}$ and $\td{}$
are no longer equal, and there are the following relations among the
Yukawa couplings:
\begin{eqnarray}
  \tu{ij}-\td{ij} &=& 9\chi_L\chi_R\st{i3}\st{j3}, \\
  \tu{ij}-\tu{ji} &=& \td{ij}-\td{ji}, \\
  \te{ij}-\te{ji} &=& 3\bigl(\td{ij}-\td{ji}\bigr), \\
  \te{ij}+\te{ji} &=& \td{ij}+\td{ji}+4(\chi_L+\chi_R)^2
  a_{i3}a_{j3}.
\end{eqnarray}
It can be easily checked that there are only 9 independent Yukawa
couplings due to these relations. In the following, we 
take $y_{33}$, $\tu{ij}$, $\st{i3}$ and $a_{i3}$ $(i=1,2)$ as
independent free parameters. All the expressions for mass eigenvalues
and mixing have been presented in the previous section. When
introducing a nonzero $\chi_R$, it can be represented by the
observable quantities as before:
\begin{eqnarray}
  \st{23}\chi_R &=& -\frac{1}{3}\left(\frac{m_s}{m_b}
    \frac{V_{\mu3}}{V_{\tau3}} +V_{cb}\right).
\end{eqnarray}
By setting $\chi_R=0$ in the above equation, the relation
(\ref{sumrule}) is reproduced. As seen in Fig.~\ref{sum}, the relation
with $\chi_R=0$ holds to very good accuracy. This fact implies that
the vacuum expectation value $\chi_R$ is significantly smaller 
than $\chi_L$. In Fig.~\ref{lr}, we display the allowed region 
for $\chi_L$ and $\chi_R$ from the masses and mixing angles for the
second and third generations alone. From this figure, one can see 
that $\chi_R$ should certainly be smaller than $\chi_L$ and may even
be zero.
When we use the central values of the masses and mixing angles, we
have $\st{23}\chi_L\sim\lambda^3$ 
and $\st{23}\chi_R\sim\lambda^4$. These two scales are interesting
because they imply, by Eq.~(\ref{78VEV}) 
and $\st{23}\sim O(\lambda^{f_2+f_3}=\lambda^2)$, that the actual VEVs
of the $\chi_L$ and $\chi_R$ components of $\phi({\bf 78})$ 
are $\langle\phi(\chi_L)\rangle \sim\lambda^3 M_P$ 
and $\langle\phi(\chi_R)\rangle\sim\lambda^4 M_P$, which are just
values around the GUT scale $10^{16}$ GeV\@. This is natural, since we
expect that the $SU(5)$ symmetry breaking is caused
by the Higgs $\phi({\bf 78})$.

\begin{figure}[htbp]
  \begin{center}
    \leavevmode
    \epsfxsize=11cm \ \epsfbox{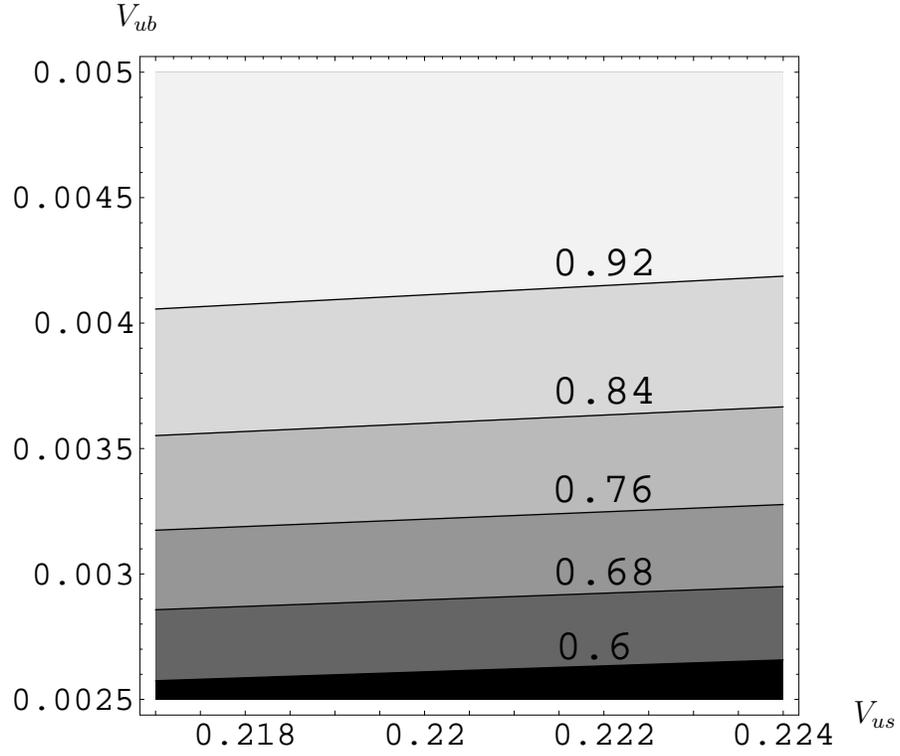}
    \put(7,20){$V_{us}$}
    \put(-272,284){$V_{ub}$}
    \caption{The prediction of the lepton 2--3 mixing angle 
      $\sin^2 2\theta_{\mu\tau}$ from the relation
      (\ref{sumrule2}). This square parameter range represents the
      experimental uncertainties of $V_{us}$ and $V_{ub}$. In this
      figure, we fix $m_s/m_b=1/40$. The predictions of the relation
      agree fairly well with the observations.}
    \label{sum2}
  \end{center}
\end{figure}

Since $\chi_R$ is responsible for the mixing angles of the first
generation (see section 5), in what follows we suppose that $\chi_R$
generally has a nonzero value. Interestingly enough, even in that case
we can find another novel relation. Among the 9 independent Yukawa
couplings, we have already fixed 4 of them, as well as the Higgs
mixing angles, from the information of the second and third
generations. The remaining 5 free parameters 
are $\tu{11}$, $\tu{12}$, $\tu{21}$, $\st{13}$, and $a_{13}$. We still
have 7 observables to be reproduced, $m_u$, $m_d$, $m_e$, $V_{us}$,
$V_{ub}$, $V_{e2}$ and $V_{e3}$. Among them, we cannot have a precise
prediction for $V_{e2}$, but we can give only an order
estimation. This is because the (right-handed) neutrino sector, which
we do not explicitly discuss in this paper, may contain more free
parameters, i.e.\ higher-dimensional couplings. We thus expect one
relation between the observables. It is found from the analytic
solutions discussed in section 5 that we now obtain another novel
relation:
\begin{eqnarray}
  \frac{V_{us}}{V_{ub}} &=& \frac{V_{\mu3}}{V_{\tau3}}\,
  \frac{m_b}{m_s}.
  \label{sumrule2}
\end{eqnarray}
This indicates that the left-handed side, which involves the first
generation, can be written only in terms of the second and third
generation parameters. Let us compare this formula (\ref{sumrule2})
with the experimental data. Figure~\ref{sum2} shows the numerical
result for Eq.~(\ref{sumrule2}), regarded as a relation between the
quark and lepton mixing angles. One can see from Fig.~\ref{sum2} that
it is also in good agreement with the experimental data.
We note that in the case $\chi_R=0$, the relation (\ref{sumrule2})
reduces to an interesting prediction between the quark mixing angles
alone,
\begin{eqnarray}
  V_{cb}V_{ub} &=& -\left(\frac{m_s}{m_b}\right)^2V_{us}.
\end{eqnarray}

For the mass eigenvalues and mixings of the first generation, we now
have the same number of parameters and observables, aside from the two
quantities $V_{e2}$ and $V_{us}/V_{ub}$ mentioned above. Thus, we can
inversely write down the parameters in terms of the observables.
\begin{eqnarray}
  \st{13} &=& -\st{23}
  \left(V-V_{e3}\,\frac{V_{\tau3}}{V_{\mu3}}\right), \\[1mm]
  \tu{12} &=& \frac{m_c}{m_t}\left(\frac{\st{13}}{\st{23}}
    +\frac{V_{ub}}{3\chi_R\st{23}V_{\tau 3}^2}\right), \\[1mm]
  \tu{21} &=& \frac{m_c}{m_t}\,\frac{\st{13}}{\st{23}}
  +\frac{m_b}{m_s}\frac{1}{V_{ub}}\frac{V_{\tau3}}{V_{\mu3}} 
  \left(\frac{m_u}{m_t} -\frac{m_d}{m_b}\frac{1}{V_{\tau3}}
    \tan\theta\right), \\[1mm]
  \tu{11} &=& \frac{m_u}{m_t} -\frac{m_c}{m_t}
  \left(\frac{\st{13}}{\st{23}} 
    +\frac{V_{ub}}{3\chi_R\st{23}V_{\tau 3}^2}\right)^2, \\[1mm]
  2(\chi_L+\chi_R)^2a_{13}^2 &=& \frac{m_\mu}{m_b}
  \frac{1}{V_{\tau3}}\,\tan\theta\left(\frac{m_e}{m_\mu}
    +\frac{1}{V_{\tau3}}\,V^2\right)
  +3V\,(\tu{12}-\tu{21})-\tu{11} \nonumber \\[1mm]
  &&~~ -\frac{V_{\mu3}}{V_{\tau3}}\,\tan\theta 
  \left(\frac{m_s}{m_b}\frac{V_{\mu3}}{V_{\tau3}} +V_{cb}\right)
  \left(V-V_{e3}\frac{V_{\tau3}}{V_{\mu3}}\right)^2.
\end{eqnarray}
On the right-handed sides of the equations, we use $\st{23}\chi_{L,R}$ 
and $\tan\theta$, which are determined by the second and third
generation data. The parameter $V$ is defined 
by $V=V_{e2}-(U^\dagger_\nu)_{12}$ and is determined by solving the
equation
\begin{eqnarray}
  V^2+aV+b &=& 0,
\end{eqnarray}
where
\begin{eqnarray}
  a &=& 2(2\alpha+\beta+\gamma), \nonumber \\[1mm]
  b &=& \frac{m_t}{m_c} \left(\frac{m_e}{m_b} \frac{1}{V_{\tau3}}
    \tan\theta+9\chi_L\chi_R\st{23}^2\alpha^2 \right)
  +(\alpha+\beta)(\alpha+\gamma) -\frac{m_u}{m_c} \nonumber\\
  &&~~ -\frac{1}{2(\chi_L+\chi_R)^2 a_{23}^2}\frac{m_t}{m_c} 
  \left(\frac{1}{V_{\tau 3}^2} \frac{m_s}{m_b}\,\alpha\tan\theta
    +\frac{m_c}{m_t}(-\beta+2\gamma)\right)^2,
\end{eqnarray}
with
\begin{eqnarray}
  \alpha &\equiv& -\frac{V_{e3}V_{\tau3}}{V_{\mu3}}, \nonumber\\[1mm]
  \beta &\equiv& \frac{-V_{ub}}{3\chi_R\st{23}V_{\tau 3}^2},
  \nonumber \\[1mm]
  \gamma &\equiv& -\frac{m_t}{m_c}\frac{m_b}{m_s}\frac{1}{V_{ub}}
  \frac{V_{\tau3}}{V_{\mu3}} \left(\frac{m_u}{m_t}
  -\frac{m_d}{m_b}\frac{1}{V_{\tau3}}\tan\theta\right).
\end{eqnarray}

As stated above, there may be subtle problems in treating such small
quantities as the masses for the first generation. Here, therefore, 
we only give a typical result for the parameters which can actually
reproduce the correct values of the observables. For the set of input
parameters
\begin{eqnarray}
  \tu{11}\sim\lambda^{5.1},\quad \tu{12}\sim\lambda^{4.8},\quad
  \tu{21}\sim\lambda^{3.9},\quad \tu{22}\sim\lambda^{3.6},\nonumber\\
  \st{23}\chi_L\sim\lambda^{2.5},\quad
  a_{23}\chi_L\sim\lambda^{2.1},\quad
  \st{13}\chi_L\sim\lambda^{3.4},\quad 
  a_{13}\chi_L\sim\lambda^{2.2},\nonumber \\
  \chi_R/\chi_L \sim0.3,\quad y_{33} \sim0.58,\quad
  \tan\theta\sim\lambda^{1.8},\quad \tan\beta\sim9.4,
  \label{para}
\end{eqnarray}
for instance, we get the following mass eigenvalues and mixings at the
GUT scale:
\begin{eqnarray}
  &&\begin{array}{lll}
    m_u \,\sim\, 1 {\rm ~MeV}, &
    m_d \,\sim\, 1 {\rm ~MeV}, &
    m_e \,\sim\, 0.3 {\rm ~MeV}, \\
    m_c \,\sim\, 0.4 {\rm ~GeV}, &
    m_s \,\sim\, 0.02 {\rm ~GeV}, &
    m_\mu\,\sim\, 0.07 {\rm ~GeV}, \\
    m_t \,\sim\, 100 {\rm ~GeV}, &
    m_b \,\sim\, 1.0 {\rm ~GeV}, &
    m_\tau\,\sim\, 1.0 {\rm ~GeV},
  \end{array} \nonumber \\[2mm]
  &&\quad\qquad V_{us} \sim0.2,\quad V_{cb} \sim0.04,\quad 
  V_{ub}\sim0.004, \nonumber \\
  &&\quad\quad V_{e2} \sim0.24+(U_\nu^\dagger)_{12},\quad
  V_{e3} \sim0.1,\quad V_{\mu3}  \sim0.7.
  \label{res}
\end{eqnarray}
Note that the parameters chosen in Eq.~(\ref{para}) are all of orders
that are consistent with our prediction $\sim\lambda^{f_i+f_j}$. Only
$\tu{11}$ and $\tu{21}$ seem larger by a factor $\lambda^1$ than our
naive expectation. This would, however, not be a large problem, since
small enhancements could occur in such combined quantities like
$T_{ij}^{u,d,e}$. Among the results in Eq.~(\ref{res}), we comment on
the lepton mixing angles. First, the 1-2 mixing cannot be determined
in the present model unless we fix the (higher-dimensional) couplings
of neutrinos. The 2-3 mixing angle is 
large ($\sin^2 2\theta_{\mu\tau}\simeq 1$), as we expect from the
generation twisting structure. In addition, the 1-3 mixing 
$V_{e3}\sim 0.1$ is consistent with the CHOOZ experimental result.

\section{Summary}
\setcounter{equation}{0}

In this paper we have examined the $E_6$ grand unified model with
E-twisted generation structure. This means that in the second and 
third generations we have taken a different choice of ${\bf 5}^*$ for
the low-energy right-handed down quarks and the left-handed charged
leptons. Such a twisting is possible, because the fundamental
representation ${\bf 27}$ in $E_6$ contains two ${\bf 5^*}$
representations of $SU(5)$. That is, in $E_6$ GUT models, we naturally
have the possibility of generation twisting without introducing extra
matter fields. We have constructed such a model, supplementing an
adjoint representation Higgs field, which is responsible for creating
differences between the quark and lepton masses and mixings (and
possibly inducing the $E_6$ gauge symmetry breaking).

Given a set of the VEVs of the adjoint Higgs components, we have
investigated the structures of fermion mass matrices which are induced
by the effective Yukawa couplings allowed by the flavor $U(1)$
symmetry. Because of the large gauge symmetry of $E_6$, we have found
that only the specific and combined Yukawa couplings can appear in the
quark and lepton mass matrices. As a consequence, we have found
several novel relations between the observables. The relations
indicate, notably, that the large lepton 2-3 mixing is related to
the precisely measured data of the quark sector alone. This is one of
the most interesting features in our $E_6$ model. As for the first
generation, the approaches utilizing flavor $U(1)$ symmetries seem to
have some disadvantages. That is, the predictability is somewhat
weakened in such case, since the relevant quantities are small, which
implies that the additional higher-dimensional operators could become
relevant unless they are forbidden by some kind of symmetries. In this
paper, we have only presented an example which can correctly reproduce
the low-energy mass eigenvalues and mixings angles.

We have shown that the generation twisting structure is naturally
incorporated in the grand unified models. In order to see whether the
model can be really viable, much more work clearly needs to be done,
for instance, on the analyses of the Higgs potential that causes the
GUT symmetry breaking and of the Majorana mass matrix structure of the 
light neutrinos. The results in this paper is encouraging enough to
motivate such efforts.

\subsection*{Acknowledgements}
The authors would like to thank Y.~Nomura, N.~Okamura and T.~Yanagida
for stimulating discussions. They also thank the organizers and
participants of Summer Institute 99 held at Yamanashi, Japan, where
this work was inspired by stimulating seminars and discussions. M.~B.,
T.~K.\ and K.~Y.\ are supported in part by the Grants-in-Aid for
Scientific Research Nos.~09640375 and 10640261, and the Grant-in-Aid
No.~9161, respectively, from the Ministry of Education, Science,
Sports and Culture, Japan.

\newpage
\appendix
\renewcommand{\theequation}{\arabic{equation}}

\section*{Appendix}

In this appendix, we present general formulas for diagonalizing 
the $3\times 3$ mass matrices. This is not only for symmetric-type
matrices but also for non-symmetric type ones, in which some
components are of the same order. Such a mass matrix appears in recent
studies on the down-quarks and lepton mass matrices which induce large
lepton mixing angles, as shown in this paper. In such a case, a
cancellation between some elements occurs and should be taken care
of. For this reason, we present formulas up to the next-to-leading
order of the expansion parameter $\lambda$. We assume that all the
matrix elements are real for simplicity.

First, consider a general symmetric matrix,
\begin{eqnarray}
  M &=& \pmatrix{
    A\lambda^6 & B\lambda^5 & C\lambda^3 \cr
    B\lambda^5 & D\lambda^4 & E\lambda^2 \cr
    C\lambda^3 & E\lambda^2 & F          \cr }.
\end{eqnarray}
It is found that this matrix has the eigenvalues
\begin{eqnarray}
  m_1 &=& {{\cal A}\over F} \left(\lambda^6 -{{\cal B}^2\over 
      {\cal D}^2F^2} \lambda^8\right), \\ 
  m_2 &=& {{\cal D}\over F} \left(\lambda^4 +{{\cal B}^2\over 
      {\cal D}^2} \lambda^6\right), \\
  m_3 &=& F+{E^2\over F}\lambda^4,
\end{eqnarray}
with
\begin{eqnarray}
  {\cal A} &\equiv& AF-C^2-{{\cal B}^2\over{\cal D}},\\
  {\cal B} &\equiv& BF-CE, \\
  {\cal C} &\equiv& BE-CD, \\
  {\cal D} &\equiv& DF-E^2.
\end{eqnarray}
The diagonalization $U M U^\dagger={\rm diag}(m_1,\,m_2,\,m_3)$ is
implemented by a unitary matrix $U$:
\begin{eqnarray}
  U &=& \pmatrix{ 
    1 & \displaystyle -{{\cal B}\over {\cal D}}\lambda 
    -{{\cal A}{\cal B}\over {\cal D}^2}\lambda^3  & 
    \displaystyle {{\cal C}\over {\cal D}}\lambda^3 
    +{{\cal A}{\cal B}E\over {\cal D}^2F}\lambda^5 \cr 
    \displaystyle {{\cal B}\over {\cal D}}\lambda 
    +{{\cal A}{\cal B}\over {\cal D}^2}\lambda^3 & 1 
    & \displaystyle -{E\over F}\lambda^2 -{{\cal B}C\over {\cal D}F}
    \lambda^4 \cr 
    \displaystyle {C\over F}\lambda^3 + {{\cal B}E\over F^3}\lambda^7
    & \displaystyle {E\over F}\lambda^2 +{{\cal D}E\over F^3}\lambda^6
    & 1 \cr} \times {\cal N},\\[3mm]
  && {\cal N} \;=\; \pmatrix{
    \displaystyle 1-\frac{1}{2}{{\cal B}^2\over {\cal D}^2}\lambda^2 & 
    0 & 0 \cr 
    0 & \displaystyle 1-\frac{1}{2}{{\cal B}^2\over {\cal D}^2}
    \lambda^2 & 0 \cr 
    0 & 0 & \displaystyle 1-\frac{1}{2}{E^2\over F^2}\lambda^4 \cr}.
\end{eqnarray}
At leading order we have 
\begin{equation}
  U \;=\; \pmatrix{
    1 & \displaystyle -{{\cal B}\over {\cal D}}\lambda&
    \displaystyle{{\cal C}\over {\cal D}}\lambda^3 \cr
    \displaystyle {{\cal B}\over {\cal D}}\lambda& 1 & 
    \displaystyle -{E\over F} \lambda^2 \cr
    \displaystyle {C\over F} \lambda^3 & 
    \displaystyle {E\over F} \lambda^2 & 1 \cr}.
\end{equation}

With this general diagonalizing formula, we can calculate the
eigenvalues and mixings for the specific types of mass matrices. The
first example is the up-type mass matrix, which is not symmetric,
although assumed to have hierarchical structure in any case:
\begin{eqnarray}
  M_u &=& \pmatrix{
    a\lambda^6 & b\lambda^5 & c\lambda^3 \cr
    d\lambda^5 & e\lambda^4 & f\lambda^2 \cr
    g\lambda^3 & h\lambda^2 & 1          \cr}.
\end{eqnarray}
We have normalized the matrix elements so that the 3-3 element becomes
1 for simplicity. We apply the above formula to the symmetric matrix
$M=M_uM_u^\dagger$. It is useful to define the following three vectors
proportional to the three rows of the matrix:
\begin{eqnarray}
  \mbox{\boldmath $c$} \equiv 
  \pmatrix{a\lambda^3 \cr b\lambda^2 \cr c \cr},\quad  
  \mbox{\boldmath $f$} \equiv 
  \pmatrix{d\lambda^3 \cr e\lambda^2 \cr f \cr},\quad 
  \mbox{\boldmath $i$} \equiv 
  \pmatrix{g\lambda^3 \cr h\lambda^2 \cr 1 \cr}.
\end{eqnarray}
We then obtain for the matrix $M=M_uM_u^\dagger$,
\begin{eqnarray}
  {\cal A} &=& {[(\mbox{\boldmath $c$}\times\mbox{\boldmath $i$})
    \times(\mbox{\boldmath $f$}\times\mbox{\boldmath $i$})]^2
    \over(\mbox{\boldmath $f$}\times\mbox{\boldmath $i$})^2} \;=\; 
  \left[a-cg-{(d-fg)(b-ch)\over e-fh}\right]^2\lambda^6,\\
  {\cal B} &=& (\mbox{\boldmath $f$}\times\mbox{\boldmath $i$})
  \cdot(\mbox{\boldmath $c$}\times\mbox{\boldmath $i$}) \;=\; 
  (e-fh)(b-ch)\lambda^4,\\
  {\cal C} &=& (\mbox{\boldmath $f$}\times\mbox{\boldmath $i$})
  \cdot(\mbox{\boldmath $c$}\times\mbox{\boldmath $f$}) \;=\; 
  (e-fh)(bf-ce)\lambda^4, \\
  {\cal D} &=& (\mbox{\boldmath $f$}\times\mbox{\boldmath $i$})^2
  \;=\; (e-fh)^2\lambda^4. 
\end{eqnarray}
At leading order, with $F=1$ in this case, the eigenvalues and mixings
can be written
\begin{equation}
  m_1^2 \;=\; {\cal A}\lambda^6,\quad
  m_2^2 \;=\; {\cal D}\lambda^4,\quad
  m_3^2 \;=\; 1,
\end{equation}
and
\begin{equation}
  U_u \;=\; \pmatrix{ 
    1 & -\displaystyle {b-ch\over e-fh}\lambda& 
    \displaystyle {bf-ce\over e-fh}\lambda^3 \cr
    \displaystyle {b-ch\over e-fh}\lambda& 1 & -f\lambda^2 \cr
    c\lambda^3 & f\lambda^2 & 1 \cr },
\end{equation}
where $U_u(M_uM_u^\dagger)U_u^\dagger
={\rm diag} (m_1^2,m_2^2,m_3^2)$.

The next example is an asymmetric-type matrix, such as the down-quark
(and the lepton) mass matrix in this paper:
\begin{eqnarray}
  M_d=\pmatrix{
    a\lambda^4 & b\lambda^3 & c\lambda^3  \cr
    d\lambda^3 & e\lambda^2 & f\lambda^2  \cr
    g\lambda& h & 1 \cr}.
\end{eqnarray}
To consider the symmetric matrix $M=M_dM_d^{\rm T}$ as before, we
define, in this case,
\begin{eqnarray}
  \mbox{\boldmath $c$} \equiv\pmatrix{a\lambda\cr b \cr c \cr},\quad
  \mbox{\boldmath $f$} \equiv\pmatrix{d\lambda\cr e \cr f \cr},\quad 
  \mbox{\boldmath $i$} \equiv\pmatrix{g\lambda\cr h \cr 1 \cr}.
\end{eqnarray}
Then we have the same expressions for ${\cal A}$, ${\cal B}$, 
${\cal C}$ and ${\cal D}$ in terms of $\mbox{\boldmath $c$}$,
$\mbox{\boldmath $f$}$ and $\mbox{\boldmath $i$}$. Of course, the
expressions are different when written with $a,\cdots,h$, but the
differences appear only in the powers of $\lambda$. Noting that 
now $F=1+h^2$ at the leading order, we can immediately write down the
eigenvalues and the mixing matrix, 
$U_d(M_dM_d^\dagger)U_d^\dagger={\rm diag}(m_1^2,m_2^2,m_3^2)$, from
the general formula
\begin{eqnarray}
  m_1^2 &=& \left[a-cg-{(d-fg)(b-ch)\over e-fh}\right]^2\lambda^8,\\
  m_2^2 &=& {(e-fh)^2\over1+h^2}\lambda^4, \\
  m_3^2 &=& 1+h^2,
\end{eqnarray}
and
\begin{equation}
  U_d \;=\; \pmatrix{ 
    1 & \displaystyle -{b-ch\over e-fh}\lambda& 
    \displaystyle {bf-ce\over e-fh}\lambda^3 \cr
    \displaystyle {b-ch\over e-fh}\lambda& 1 & 
    \displaystyle -{f+eh\over1+h^2}\lambda^2 \cr
    \displaystyle {c+bh\over1+h^2}\lambda^3 & 
    \displaystyle {f+eh\over1+h^2}\lambda^2 & 1 \cr}.
\end{equation}

Finally, we consider the charged-lepton mass matrix, the transpose of
which takes the same form as the down-quark one:
\begin{eqnarray}
  M_e^{\rm T} &=& \pmatrix{
    a\lambda^4 & b\lambda^3 & c\lambda^3  \cr
    d\lambda^3 & e\lambda^2 & f\lambda^2  \cr
    g\lambda& h & 1 \cr} =M_d.
\end{eqnarray}
Then, clearly the eigenvalues $m_1,\,m_2,\,m_3\,(>0)$ are the same as
those for the down-quark matrix $M_d$. The mixing matrix realizing 
$U_e(M_eM_e^\dagger)U_e^\dagger={\rm diag} (m_1^2,m_2^2,m_3^2)$ is
found from the relation 
$U_dM_dU_e^\dagger=D_m\equiv{\rm diag}(m_1,m_2,m_3)$ as
\begin{eqnarray}
  U_e &=& D_m^{-1} U_d M_e^\dagger \nonumber \\
  &=& \pmatrix{ 
    1 & \displaystyle -{d-fg\over e-fh}\lambda&
    \displaystyle -{eg-dh\over e-fh}\lambda\cr
    \displaystyle {d(1+h^2)-g(f+eh)\over\sqrt{1+h^2}}\lambda& 
    \displaystyle {1\over\sqrt{1+h^2}} & 
    \displaystyle -{h\over\sqrt{1+h^2}} \cr
    \displaystyle {g\over\sqrt{1+h^2}}\lambda& 
    \displaystyle {h\over\sqrt{1+h^2}} & 
    \displaystyle {1\over\sqrt{1+h^2}} \cr}.
\end{eqnarray}
{}From the form of this mixing matrix, we can see that it depends only
on the parameter $h$ whether the large mixing angle between the second 
and third generations is realized or not.


\end{document}